\documentclass[a4paper,11pt]{article}
\usepackage{pos}

\title{Robust extraction of power corrections and nuclear dynamics from DIS at large x }

\ShortTitle{Robust power corrections extraction}

\author*[a,b]{Alberto Accardi}
\emailAdd{alberto.accardi@cnu.edu}
\affiliation[a]{Christopher Newport University \\ Newport News, Virginia 23606, USA}
\affiliation[b]{Jefferson Lab \\ Newport News, Virginia 23606, USA}
\affiliation[c]{IRFU, CEA, Universit\'e Paris-Saclay, F-91191 Gif-sur-Yvette, France}

\author[c]{Matteo Cerutti}
\emailAdd{matteo.cerutti@cea.fr}

\abstract{We present recent updates from the CTEQ-JLab (CJ) global PDF analysis, focusing on the interplay and implementation systematics of the HT and offshell correction (\texttt{CJ22ht}). We also discuss preliminary results of the \texttt{CJ25} global analysis, showing the impact of the full JLab 6 GeV datasets, that we recently collected in a comprehensive DIS database, and having a first look at early JLab 12 GeV measurements. We finally offer a few thoughts on how future data may help unraveling the nuclear and partonic dynamics in light nuclei.}

\FullConference{XXXII International Workshop on Deep Inelastic Scattering and Related Subjects (DIS2025)\\
24-28 March, 2025\\
Cape Town, South Africa\\}

\renewcommand{\hookAfterAbstract}{%
\vskip1cm\par\bigskip
\textsc{JLab preprint number}: JLAB-THY-25-4602
}

\begin{document}
\maketitle

\section{Introduction}

%
The CTEQ-JLab (CJ) collaboration on global QCD analysis~\cite{Accardi:2016qay,Accardi:2023gyr,Li:2023yda,Cerutti:2025yji} aims at extracting flavor-separated PDFs at large $x$ by combining diverse datasets. Proton Deep-Inelastic Scattering (DIS) data constrain the $u$-quark distribution, while the $d$-quark relies heavily on deuteron-target DIS, $W$-boson asymmetries, and tagged DIS from BoNUS at JLab. However, a large fraction of the DIS data lies at low or intermediate photon four-momentum squared $Q^2$, where nuclear corrections and power-suppressed $1/Q^2$ effects become significant. These include target mass corrections (TMCs), which are calculable albeit with theoretical uncertainty, higher-twist (HT) contributions from multiparton correlations and other nonperturbative effects. HT terms are typically fitted, and their modeling can impact PDF extractions. Furthermore, with little data available at large $Q^2$, the leading-twist nuclear corrections to deuteron DIS calculation -- in particular, offshell nucleon deformations -- correlate with the extraction of the HT terms due to their ability to modify the $d$-quark result at large $x$.

In this contribution, we summarize the \texttt{CJ22ht} global QCD analysis~\cite{Cerutti:2025yji} and show that an isospin-independent HT implementations can strongly bias the large-$x$ behavior of the $F_2^n/F_2^p$ ratio, which strongly correlates with the off-shell function. We identify a more robust HT implementation and propose experimental observables to better isolate the off-shell effects.

We also present preliminary results of the \texttt{CJ25} analysis, that includes for the first time the full JLab 6 inclusive DIS dataset (see the DIS database described in Ref.~\cite{Li:2023yda}). These data further constrain the nucleon's large-$x$ structure and reduce model dependence in nuclear corrections. Finally, we discuss the impact of early JLab 12 GeV data, which will further sharpen our understanding of partonic dynamics in light nuclei. 

\section{Interplay of higher-twist and off-shell corrections}
\label{s:HTvsOS}


\vspace{10pt}
\noindent
\textit{Off-shell corrections}. The partonic structure of a bound nucleon  depends on its four-momentum squared $p^2$ with non-negligible deformations at large Bjorken-$x$. In the weakly-bound deuteron, one can Taylor-expand the nucleon structure around $p^2 \simeq M^2$ (for more details, see Refs~\cite{Kulagin:1994cj,Kulagin:1994fz}). We perform this expansion at the PDF level,
\begin{equation}
\phi(x,Q^2,p^2) = \phi^{\text{free}}(x,Q^2) \bigg ( 1 + \frac{p^2 - M^2}{M^2} \delta f (x) \bigg ) \, ,
\label{e:off_pdf}
\end{equation}
where $\phi^{\text{free}}$ is the free-nucleon PDFs, and the $\delta f$ ``off-shell function'' quantify the deformation of a parton belonging to bound nucleon. Since $\delta f$ is defined as a ratio of PDFs, we assume it to be independent of $Q^2$. We assume it to be flavor independent since we only include deuteron target data. We parametrize and fit the off-shell functions with a generic 2$^{\rm nd}$ order polynomial in $x$,
$
    \delta f (x) = \sum_{n=0}^{2} a_{\text{off}}^{(n)} x^n.
    \label{e:off_poly}
$
We did not see any significant benefit of using polynomials of higher degrees, except as a tool to identify $x \lesssim 0.6$ as the momentum fraction region in which the \text{CJ22ht} dataset can constrain $\delta f$. The full JLab data will likely warrant the use of 3rd degree polynomials.

\vspace{10pt}
\noindent
\textit{Higher-twist corrections}. 
The $1/Q^2$ corrections to the leading-twist (LT) expression of DIS observables have different nature (see Ref.~\cite{Cerutti:2025yji} for a detailed discussion) and mainly affect the large-$x$ region. The combination of all these sources is commonly referred to as ``higher-twist'' corrections even though they do not only arise from twist-4 hadron matrix elements.
Including suitable power correction terms in a global QCD fit allows one to reliably extract the leading-twist dynamics (PDFs and off-shell function) of the nucleon. However, a study of the impact of the extracted HT function on correlated quantities such as the $d/u$ ratio and the off-shell deformation $\delta f$ is needed.

Power corrections can be implemented in global QCD analyses by means of \textit{multiplicative} or \textit{additive} modifications of the target-mass-corrected (TMC) structure functions, 
\begin{align}
    F_{2N}^\text{mult}(x,Q^2) &= F_{2N}^\text{TMC}(x,Q^2) \bigg ( 1 + \frac{C(x)}{Q^2} \bigg ) \ , 
    \label{e:ht_mult}
    \\
      F_{2N}^\text{add}(x,Q^2) &= F_{2N}^\text{TMC}(x,Q^2) + \frac{H(x)}{Q^2} \ , 
    \label{e:ht_add}
\end{align}
and fitting the $x$-dependent ``higher-twist'' functions $C(x)$ or $H(x)$ for a nucleon $N=p,n$.

There is no strong a priori reason for choosing either implementation, but any choice implicitly introduces consequences on the $Q^2$ scaling and isospin dependence of the HT functions that are usually ignored. For instance, one can rewrite multiplicative implementation~\eqref{e:ht_mult} in an additive way,
\begin{equation}
    F_{2N}^\text{mult}(x,Q^2) = F_{2N}^\text{TMC}(x,Q^2) + \frac{\tilde{H}_N(x,Q^2)}{Q^2} \ .
    \label{e:ht_tilde}
\end{equation}
However, the resulting additive $\tilde H$ coefficient inherits both isospin dependence and $Q^2$ evolution from the $F_2$ structure function, namely, 
$
    \tilde{H}_N(x,Q^2) = F_{2N}(x,Q^2) C(x) \ .
\label{e:Htilde}
$

\ \\
\noindent
\textit{Identification of their interplay}. The ratio of neutron to proton DIS structure functions can be determined from proton and deuteron target DIS data after removing nuclear corrections. However, the presence of both HT and off-shell corrections affect the determination of the deuteron structure function, and biases in either one can be compensated by the fitted parameters of the other. Consequently, the  $n/p$ ratios extracted with a multiplicative or an additive HT implementation may end up to be incompatible with each other~\cite{Cerutti:2025yji}. The fit may then compensate this effect and bias the extraction of leading twist quantities such as the $d/u$ ratio and the off-shell deformation function $\delta f$.

For example, if we consider isospin-independent \textit{multiplicative} HT functions, $C(x) \equiv C_p(x) = C_n(x)$, the correction simply cancels in the $n/p$ ratio and one obtains the same limit at large $x\rightarrow1$ as in a LT calculation: $n/p \simeq 1/4$.
With isospin-independent \textit{additive} HT corrections, $H(x) \equiv H_p(x) = H_n(x)$, we obtain instead $n/p \simeq 1/4 + \Delta$, with $\Delta =  \frac{27}{16} \frac{H/u}{Q^2}$. Therefore, a larger tail originates with additive HT than the multiplicative HT. This is a direct consequence of the implementation choice, potentially leading to an overestimate of the $n/p$ ratio.

\begin{figure}[tbh]
\centering
\includegraphics[width=0.97\textwidth]{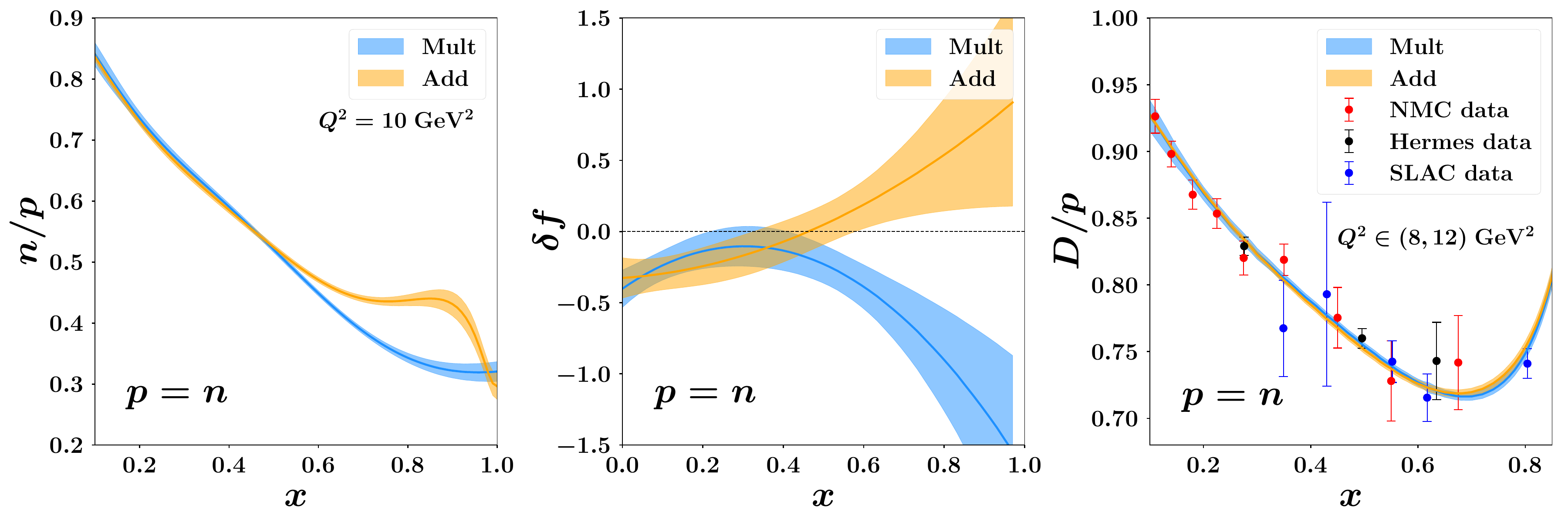}
\vskip-0.2cm
\caption{
Results of the \texttt{CJ22} analyses with isospin-independent ($p=n$) additive (orange band) or multiplicative (blue band) HT corrections. \textit{Left panel}: $n/p$ ratio of $F_2$ structure functions at $Q^2=10$ GeV$^2$. \textit{Central panel:} offshell function. \textit{Right panel}: $D/p$ ratio at $Q^2=10$ GeV$^2$ compared to a selection of experimental data.
Bands represent $T^2=2.7$ uncertainties (see Ref.~\cite{Accardi:2015lcq} for more details).}
\label{f:Res_HTindep}
\end{figure}

We prove this bias in Fig.~\ref{f:Res_HTindep}, which shows results of the CJ global QCD fit performed with isospin-independent HT corrections. While the experimental data on the $D/p$ ratio are described in the same way, the unobserved $n/p$ ratio extracted in the two implementations strongly deviate from each other at $x\gtrsim 0.6$, with the additive HT curve much larger than the multiplicative one at $x\simeq 0.85$. At even larger $x$ values, there is little data available to constrain the fit, and the additive $n/p$ ratio steeply dives down to zero, which is a built-in behavior in the additive implementation since $H(x) \to 0$ as $x \to 1$. (In fact, we may think about this as an additional parametrization bias arising in the additive implementation.)
At the same time, the $d/u$ ratio (not shown here) does not display a dependence on the HT implementation model. The behavior of the $n/p$ ratio is instead correlated with the extracted off-shell function at $x\gtrsim 0.5$. Being large and positive in the additive implementation while large and negative in the multiplicative one, it effectively compensates the HT-dominated $n/p$ behavior in the fit to experimental data.

\begin{figure}[t]
\centering
\includegraphics[width=0.97\textwidth]{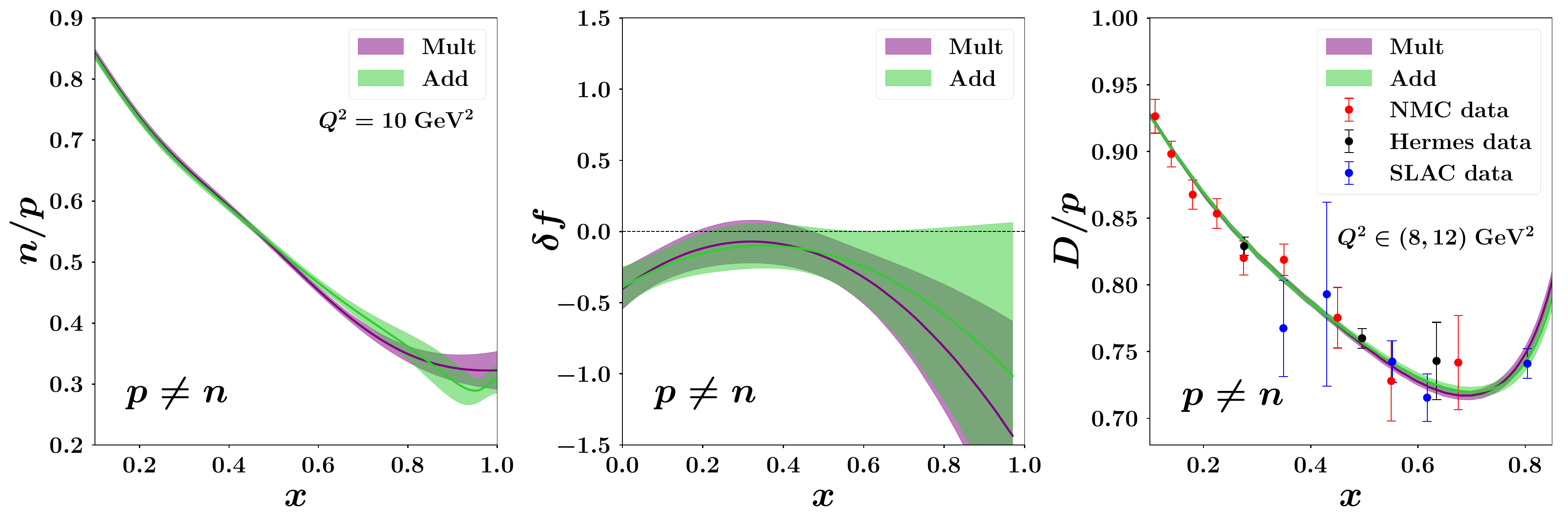}
\vskip-0.2cm
\caption{
Comparison of the isospin-dependent ($p\neq n$) additive (green band) or multiplicative (purple band) implementation of the HT corrections in the CJ global analysis. Other details as in Fig.~\ref{f:Res_HTindep}.}
\vskip-0.2cm
\label{f:Res_HTdep}
\end{figure}

The HT implementation bias we have identified can be removed by considering isospin-dependent HT terms. Indeed, for additive HT ($p\neq n$) one obtains
$n/p \simeq \frac{1}{4} + \Delta'$, with $ \Delta' = \frac{9}{16} \frac{{H}_p/u}{Q^2}$ when assuming $H_n \approx \frac12 H_p$ at large $x$~\cite{Alekhin:2003qq}. 
With the bias removed, we expect to obtain more stability in the extracted quantities (offshell function and $n/p$ ratio).
This is precisely the results of our fits with with isospin-independent HT corrections reported in Fig.~\ref{f:Res_HTdep}: the $n/p$ ratios and the extracted $\delta f$ functions are now largely independent of the choice of HT implementation. The remaining differences can be considered a phenomenological systematic uncertainty in the fit. 

\section{New insights from JLab data}

Having established a stable and systematically controlled implementation of higher-twist (HT) and off-shell corrections, we are now in a position to investigate additional experimental constraints from the Jefferson Lab 6 GeV and 12 GeV programs. 
To this purpose, we make use of the comprehensive database of inclusive DIS measurements from world-wide experiments developed in Ref.~\cite{Li:2023yda}. Jefferson Lab data complement those from SLAC, NMC and BCDMS and offer good coverage in the low-$Q^2$ and moderate-to-large $x$ region. This makes them especially suitable for obtaining insights on both nuclear and nonperturbative QCD effects. 

We started from a baseline fit that excludes the E00-116 $F_2$ proton and deuteron data already considered in \texttt{CJ22ht}~\cite{Cerutti:2025yji}. To simplify the discussion for this contribution, we left the tagged $n/d$ DIS data from the BoNUS experiment in the baseline fit, and assess here the impact of JLab 6 inclusive measurements only (proton and deuteron target data from the E06-009, E94-110, E03-103, E99-118, JLCEE96, and E00-116). Likewise, we will only show results for the multiplicative HT implementation, which do not differ much from the additive implementation when considering isospin-dependent corrections.
After a careful analysis of the low invariant mass region, we increased the $W^2$ cut from 3~GeV$^2$ (used in the CJ22 fits) to 3.5~GeV$^2$ to better exclude resonant electron scattering.

The baseline fit includes a total of 4019 data points that are fitted by 29 free parameters, leading to a very nice $\chi^2/N_{\rm dat}$ of 1.04. When we introduce the JLab 6 GeV data discussed above, the total number of data points increases to 4728 and the $\chi^2/N_{\rm dat}$ decreases to 1.02. While a detailed analysis is still underway, this initial result is a promising indication that there are no significant tensions between the baseline dataset and the newly introduced JLab 6 inclusive data, and that the theoretical framework on which our analysis is based nicely accommodates them.

\begin{figure}[htb]
\centering
\includegraphics[width=1.0\textwidth]{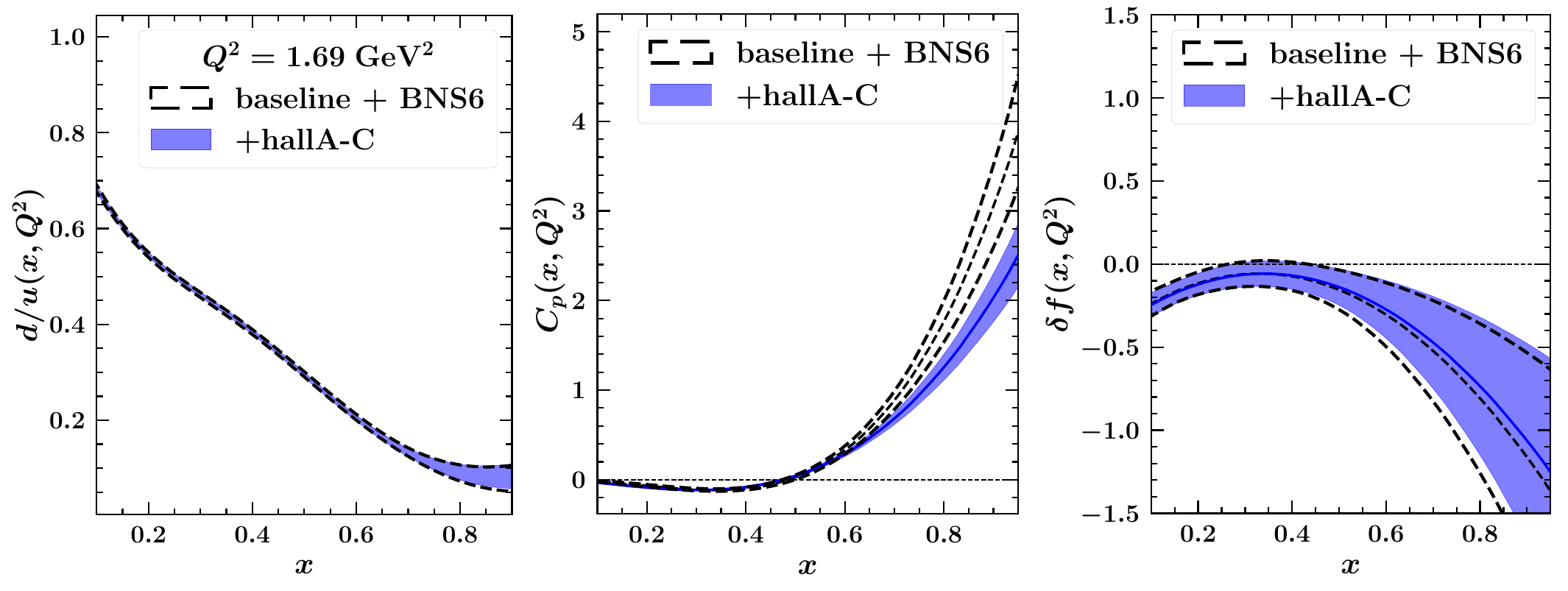}
\vskip-0.2cm
\caption{Comparison between extracted quantities from baseline fit (dashed empty bands) and fit including JLab 6 data (blue bands). Left panel: $d/u$ PDF ratio at $Q^2=Q_0^2=1.69$ GeV$^2$. Central panel: proton HT function $C_p$; right panel: offshell function $\delta f$. Error bands account for $T^2=2.7$ uncertainties.
}
\vskip-0.2cm
\label{f:Res_JLab}
\end{figure}

In Fig.~\ref{f:Res_JLab}, we show the comparison between a selection of the extracted quantities from our fit including JLab 6 data (blue bands) and from the baseline fit (dashed empty bands). In the left panel, we display the $d/u$ PDF ratio at $Q^2=Q_0^2=1.69$ GeV$^2$; in the central panel the proton HT function $C_p$; in the right panel the offshell function $\delta f$. Error bands account for $T^2=2.7$ uncertainties.
We observe that the introduction of JLab 6 GeV measurement provides a significant impact on the extracted proton HT function $C_p(x)$ on both its size and uncertainties at large-$x$. The neutron HT function $C_n(x)$ (not shown) displays a similar  uncertainty reduction but its central value is not modified. In the left and right plot we see that the impact on the extracted $d/u$ ratio and the offshell function $\delta f$, respectively, is small. Overall, JLab 6 data increase the value of the extracted $n/p$ ratio at large $x$.

A small impact on leading twist quantities is expected because the newly introduced JLab data cover only a small $Q^2$ range, and are thus more sensitive to the higher-twist corrections. Indeed the leading-twist $d/u$ remains mainly constrained by the $W$-boson asymmetry Tevatron datasets, and modifications of the off-shell function are difficult to detect when only low-$Q^2$ data are considered. 

With additional measurements at higher $Q^2$ we could instead expect the offshell function to receive more constraints and decorrelate from the HT functions. Our preliminary studies have shown that the recently released data on the $D/p$ ratio from Hall C \cite{Biswas:2024diw}, among other DIS measurements using the 12 GeV beam at Jefferson Lab, are essential to obtain a robust extraction of the off-shell and HT corrections in global fits, and provide a sizeable impact on the extracted $d/u$ PDF ratio and offshell function. The high-precision, proton-tagged DIS measurements from BONuS12 \cite{BONuS12} which are directly sensitive to the neutron structure, would then allow us to further investigate the dynamics of off-shell nucleon deformations.



\section*{Acknowledgments}

\noindent
We are grateful to our CTEQ-JLab colleagues for many discussions, and for their collaboration on the results presented in this contribution. This work was supported in part by the  U.S. Department of Energy (DOE) contract DE-AC05-06OR23177, under which Jefferson Science Associates LLC manages and operates Jefferson Lab, and by DOE contract DE-SC0025004.

\bibliography{DIS25biblio}
\bibliographystyle{myrevtex}

\end{document}